\documentclass{aa}  
\usepackage[latin1]{inputenc}
\usepackage{graphicx}
\usepackage{longtable}
\usepackage{color}
\usepackage{url}
\usepackage[breaklinks=true,pagebackref=false]{hyperref}
\usepackage{multirow}
\usepackage[table]{xcolor}
\usepackage{txfonts}
\usepackage{natbib}
\usepackage{colortbl}
\bibpunct{(}{)}{;}{a}{}{,} 

\definecolor{gris}{gray}{0.9}

\begin{document}
\title{Statistical and Numerical Study of Asteroid Orbital Uncertainty
}

   \author{J. Desmars
          \inst{1,2}
          \and
          D. Bancelin\inst{2}
          \and
          D. Hestroffer\inst{2}
          \and
          W. Thuillot\inst{2}
          }

 \offprints{J.Desmars, desmars@imcce.fr}

   \institute{Shanghai Astronomical Observatory, 80 Nandan Road, The Chinese Academy of Science, Shanghai, 200030, PR China
         \and
             IMCCE - Observatoire de Paris, UPMC, UMR 8028 CNRS, 77 avenue Denfert-Rochereau,
		75014 Paris, France\\
              \email{[desmars,bancelin,hestro,thuillot]@imcce.fr}
             }


 
  \abstract 
   {The knowledge of the orbit or the ephemeris uncertainty of asteroid presents a particular interest for various purposes. These quantities are for instance useful for recovering asteroids, for identifying lost asteroids or for planning stellar occultation campaigns. They are also needed to estimate the close approach of Near-Earth asteroids, and subsequent risk of collision. Ephemeris accuracy can also be used for instrument calibration purposes or for scientific applications. }
   {Asteroid databases provide information about the uncertainty of the orbits allowing the measure of the quality of an orbit. The aims of this paper is to analyse these different uncertainty parameters  and to estimate the impact of the different measurements on the uncertainty of orbits.}
   {We particularly deal with two main databases ASTORB and MPCORB providing uncertainty parameters for asteroid orbits. Statistical methods are used in order to estimate orbital uncertainty and compare with parameters from databases. Simulations are also generated to deal with specific measurements such as future Gaia or present radar measurements.}
   {Relations between the uncertainty parameter and the characteristics of the asteroid (orbital arc, absolute magnitude, ...) are highlighted. Moreover, a review of the different measuments are compiled and the impact of these measures on the accuracy of the orbit is also estimated.}
   {}


   \maketitle
%

\section{Introduction}

Asteroids, since the time of their first discoveries in the nineteenth century, have raised the need to compute and to predict their position for ensuring subsequent detections and observations. While methods for computing orbits or predicting positions of planets and comets had been developed at that time, the case of asteroids (or minor planets) raised new challenges; because no specific assumption can be made on the orbital elements, that are all completely unknown. This was successfully solved by Gau{\ss} \citep{Gauss1809, Gauss1864} and his method of orbit computation from the knowledge of three topocentric positions. That method, presented as ``a judicious balancing of geometrical and dynamical concepts'' by \citet{Escobal1965}, proved to be remarkably efficient providing---shortly after the discovery of Ceres by Piazzi, in 1801, and observations over only a 40 days covering a 3\,degree arc---a prediction for the next apparition of Ceres twelve month later, to better than 0.1 degree. Without being exhaustive, one can also mention here, for completion, the method of Laplace \citep[e.g.][]{Poincare1906}, based on the knowledge of a position and its derivatives, which shows several similarities on the fundamental scheme \citep{Tisserand1899,Celletti2005}. These methods \citep[and reference therein]{Herget1948,Dubyago1961,Escobal1965,Roy2005,Milani2010} paved the way for many more discovery of minor planets, or asteroids, and continuous observations of these bodies. The number of known asteroids is still increasing, presently at the end 2012 there are slightly less than 600,000 catalogued bodies in the {\sf astorb} database \citep{astorb}; and it is interesting to remind the remark extracted from the preface of E. Dubois, the French translator of Gauss' work \citep{Gauss1864}: \newline
``Or il est bien probable que la zone situ\'ee entre Mars et Jupiter n'est pas encore suffisamment explor\'ee et que le chiffre de 79 auquel on est arriv\'e, sera encore augment\'e. Qui sait ce que r\'eserve l'avenir !!... Bient\^{o}t alors les astronomes officiels n'y pourront plus suffire, si des calculateurs d\'evou\'es \`a l'astronomie et \`a ses progr\`es ne leur viennent aussi en aide de ce cot\'e\footnote{``It is likely that the zone between Mars and Jupiter is not yet sufficiently explored and that the number of 79 that has been reached will still be increased. Who knows what the future will hold for us!! Soon official astronomers will not suffice, if some calculators devoted to astronomy and its progress do not come to their rescue also on that side.'' It is clear that two centuries later modern electronic computing machines came to the rescue and are now impossible to circumvent.}''. Preliminary orbit computations are thus needed to provide some initial knowledge of the celestial object's orbit. They differ in nature, as an inverse problem, from the direct problem of subsequent position prediction and orbit propagation that can be more complete in terms of correction to the observations and force models. 

Nowadays, ephemerides are commonly needed for various practical or scientific use: to prepare an observation program or space mission rendez-vous, to be able to cross-match a source or to identify a known asteroid observed in a CCD frame, to link two observed arcs as being of the same object, for predicting stellar occultations, for developing planetary ephemerides, for testing new dynamical models and physics, for contributing to local tests of the general relativity, for deriving asteroid masses from close encounters, or non-gravitationnal parameters, etc. 
During previous centuries these objects were considered as planets or minor planets, and analytical or semi-numerical theories could be applied to each body before the advent of modern powerful computers \citep[e.g.][]{Leveau1880,Brouwer1937,Morando1965}. Ephemerides are now computed directly by numerical integration of the equation of motions for the orbit propagation using different integrators. Such numerical integration requires the use of a standard dynamical model and associated parameters for all solutions to be consistent, and the knowledge of initial conditions to the Cauchy problem as given by orbital parameters at a reference epoch. The latter are obtained, as for any solar system object, by a fit to the observations and are updated regularly. The terminology 'ephemerides' is broadly used for any quantity related to the computation of such orbit propagation of a celestial object, giving its position and motion (spherical cartesian, apparent, astrometric, as seen from various centre, in various reference planes, at various dates, on different time scales, etc.). 

In addition to these calculated values from orbit propagation and subsequent transformations corresponding generally to a nominal solution (e.g. least-squares with a standard dynamical model), knowledge of the ephemeris uncertainty, precision and accuracy is often mandatory as is the case for any physical quantities. It is also often useful to know the degree of confidence one can have in a predicted quantity. Further, the ephemerides uncertainty is needed in the various fields presented above and in particular for planning observations using instruments with small field of view, for instrument calibration purposes, or when one needs to quantify the probability of an impact of a Near-Earth Object (NEO) with a terrestrial planet, or for efficient planning of stellar occultation campaigns (in particular if it involves large number of participants and/or large telescopes), etc.

The precision of an ephemeris can be incorrectly reduced to the precision of the observations it is based on. Of course, all things being equal, the higher the precision of the observation the better the theory and ephemerides; but this will not yield an indication of how fast the precision for any predicted or extrapolated quantity is being degraded. Indeed, the precision of the ephemerides is a quantity varying with time; when extrapolating the motion to dates far into the future or in the past (several centuries), the precision will globally be poorer; up to the point where chaotic orbits on time span of million years make such ephemeris prediction unrealistic.
An ephemeris uncertainty can be decreased on a short time span---for instance if required by a flyby or space mission to a solar system body---with last minutes observations. However, if no more observations are added in the fitting process, the ephemerides precision will inevitably degrade in time. 
The accuracy of an ephemeris as the precision of the theory for the dynamical model and its representation can be internal or external. Internal precision refers to the numerical model used (terms of the development in an analytical theory or machine precision in case of numerical integration of the ODE for the equations of motions) or additionally to the representation used to compress such ephemerides (Chebychev polynomials, mixed functions, etc.). External precision refers to the adjustment to the observations and hence to the stochastic and systematic errors involved in the measurements and data reduction, in the validity of the models used to fit and transform the data, in the uncertainty in the considered parameters, etc. We will deal in the following with the external errors, which accounts for most of the uncertainty in present ephemerides of asteroids and other dwarf planets, and small bodies of the Solar System.

All ephemerides require observations and measurents related to the dynamical model of the equations of motion. Current ground-based surveys (LINEAR, Catalina, Spacewatch, C9) which where basically designed to detect 90\% of the largest NEOs provide most of such data; they are completed by some scientific programs such as radar observations from Arecibo and Goldstone for NEOs, the CFHTLS Ecliptic Survey for the Trans-Neptunian Objects \citep{Jones2006} or the Deep Ecliptic Survey (DES) for Kuiper Belt Objects and Centaurs \citep{Elliot2005}. In the very near future surveys such as Pan-starrs, Gaia, LSST, will also provide large amount of astrometric positions. 

In this paper, we investigate the orbital uncertainty of asteroids in the numerical and statistical way. First, we briefly present the two main databases of orbital parameters of asteroids. In particular, we present and compare uncertainty parameters provided by these databases (Sect.~\ref{S:uncertainty}). 

In Sect.~\ref{S:sec_ARC_CEU} we highlight relations between orbital uncertainty and other asteroid parameters such as dynamical classes of asteroids, magnitude. We specifically study the relation between orbital arc and uncertainty, in particular for short orbital arcs. Eventually, in Sect.~\ref{S:measurements} we present statistical information about astrometric measurements of asteroids and we quantify the impact of astrometric measurements in the radar and Gaia space mission context.

\section{Ephemeris uncertainty parameters}
\label{S:uncertainty}
The number of discovered asteroids has exceeded 590,000 at the beginning of October 2012, and the discovery rate is still of about 1800 new asteroids per month\footnote{Average of the first half of year 2012.}. Currently, four main centres provide asteroid orbital databases (Minor Planet Center, Lowell Observatory, Jet Propulsion Laboratory, Pisa University). In this paper, we mainly deal with two of these databases: \textsf{astorb} from Lowell Observatory  \citep{astorb} and \textsf{mpcorb} from \cite{mpcorb}. A total of 590,095 asteroids are compiled in \textsf{astorb} and 590,073 in \textsf{mpcorb} as from October 5th 2012.
The two databases provide similar asteroid parameters, in particular: 
\begin{itemize}
\item \textbf{name/number}: name or preliminary designation, asteroid number;
\item \textbf{osculating elements}: semi-major axis $a$, eccentricity $e$, inclination $I$, mean anomaly $M$, argument of perihelion $\omega$, longitude of ascending node $\Omega$ at reference epoch;
\item \textbf{magnitude}: absolute magnitude $H$, slope parameter $G$; 
\item \textbf{observations}: number of astrometric observations, arc length or year of first and last observation.
\end{itemize}

The two databases also provide parameters about predictability of the ephemerides. Only one for \textsf{mpcorb} which is the U parameter and five parameters for \textsf{astorb} which are CEU, PEU, next PEU and the two greatest PEU.\\

The \emph{uncertainty parameter} U is an integer value between 0 and 9, where 0 indicates a very small uncertainty and 9 an extremely large uncertainty. Detailed information about the computation is given in \cite{MPC_U}. Briefly, the U parameter is computed thanks to another parameter, $RUNOFF$, which depends on the uncertainty in the time of passage at perihelion, the orbital period and its uncertainty. $RUNOFF$ expresses the uncertainty in longitude in seconds of arc per decade. The U parameter is derived by a logarithmic relation of $RUNOFF$.
The quality of the orbit can be quickly determined with the uncertainty parameter U.\\ 

The \textsf{astorb} database provides five parameters related to the ephemeris uncertainty. In our study, we specifically dealt with two of them, \emph{current ephemeris uncertainty} (CEU) and the \emph{rate of change of CEU}. The CEU matches to the \emph{sky-plane uncertainty} $\sigma_\psi$ at a date\footnote{The date of CEU is usually from 0 to 40 days before the epoch of osculation depending on the update of the database. In our case, the date of CEU and the epoch of osculation are the same September 30th 2012. For previous updates, the difference can reach about 45 days} provided by \textsf{astorb}. A brief description of the uncertainty-analysis technique is presented in \citet{Y87} and all the details can be found in \citet{MB93}. The orbit determination provides the covariance matrix of the orbital elements $\Lambda$. Linear transformations then give the covariance matrix $\Sigma$ in spherical coordinates (in right ascension and declination):

\begin{equation}
\Sigma = \Psi \Lambda \Psi^{T} 
\end{equation}

where $\Psi$ is the matrix of partial derivatives between spherical coordinates and orbital elements. Finally, by propagating this covariance matrix at a given date (date of CEU), the 
sky-plane uncertainty can be determined as the trace of the matrix \citep{MB93}. The CEU is determined in two body linear approximation \citep{astorb}. 
The rate of change of CEU (noted CEU rate) is in arcsec/day. According to \cite{MB93}, in linear approximation, uncertainties in spherical coordinates increase linearly with time in the two body approximation.

\subsection{Comparison of the ephemeris uncertainty parameters}

\begin{figure}[h!] 
\centering 
\includegraphics[width=\columnwidth]{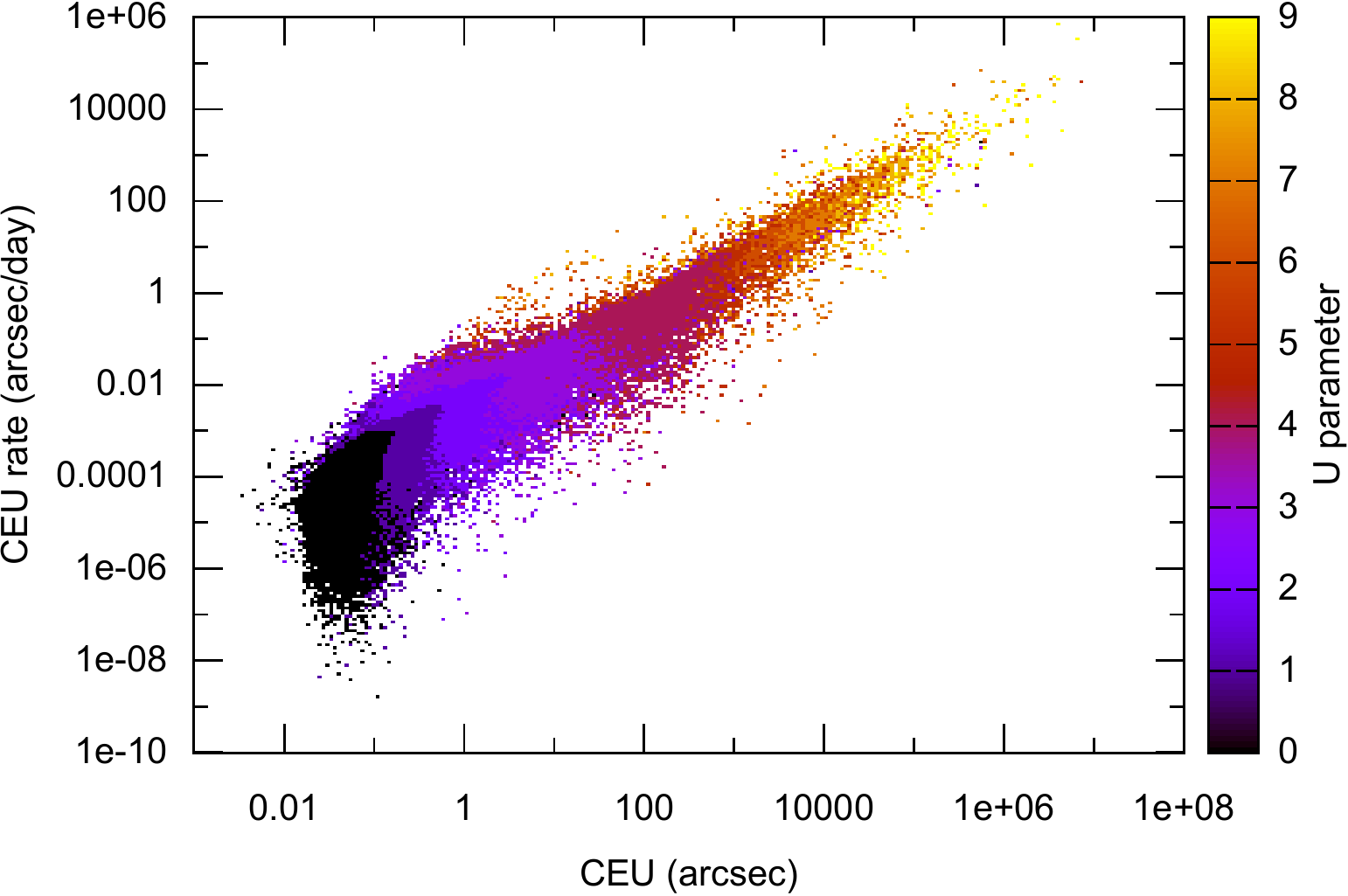}
\caption{CEU, CEU rate, U parameter for asteroids in common to the \textsf{astorb} and \textsf{mpcorb} databases.}\label{fig_ceu_ceup_u}
\end{figure}

The U parameter can be compared to the CEU and its rate of change. Figure \ref{fig_ceu_ceup_u} shows a correlation between these parameters. Qualitatively and as expected, asteroids with a low U parameter have also a low CEU and a CEU rate. 

The measure of the uncertainty could be provided by only one of these parameters. Nevertheless, those two parameters are not perfect and some problems can appear.  

\begin{figure}[h!] 
\centering 
\includegraphics[width=\columnwidth]{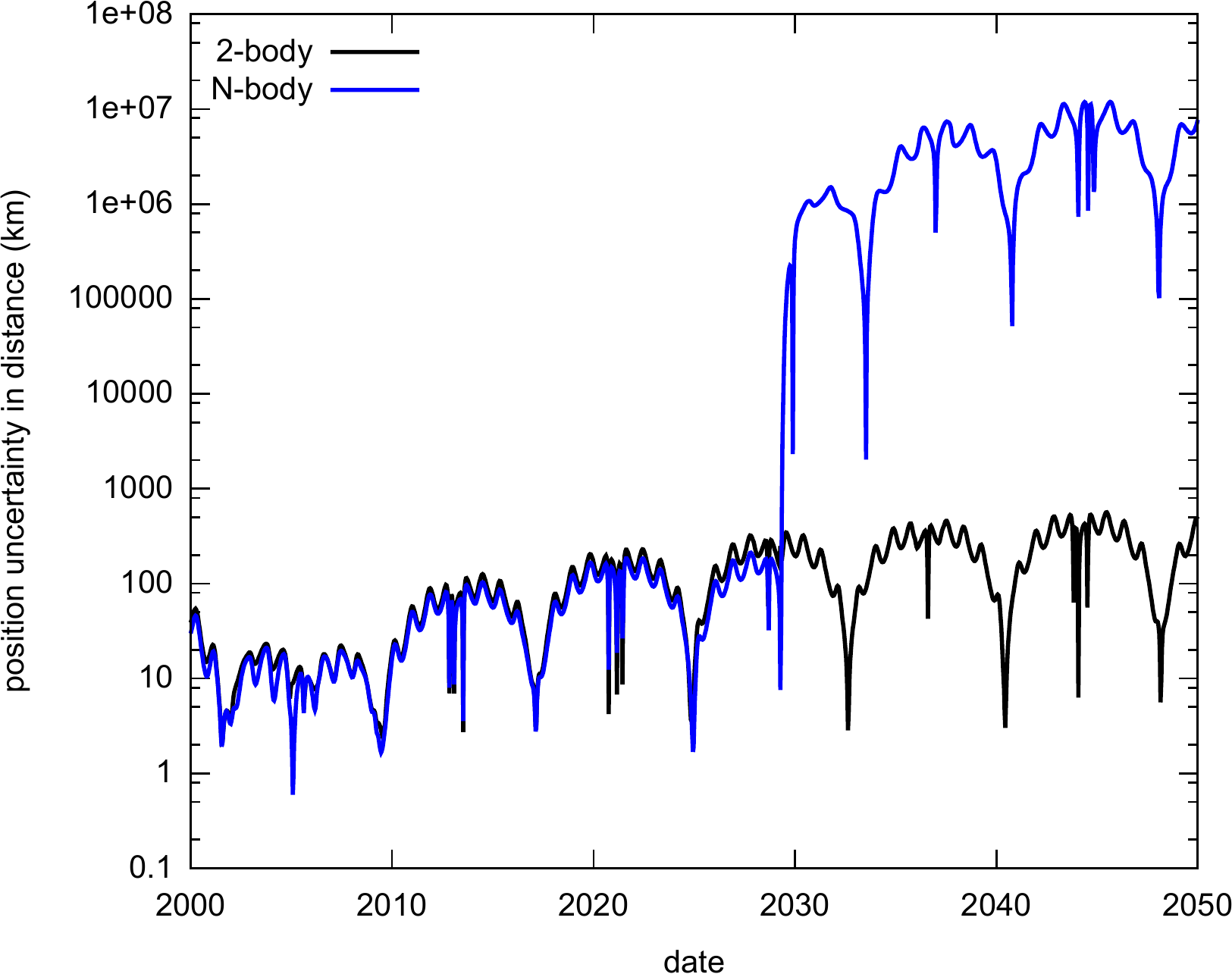}
\caption{Orbital uncertainty in distance of asteroid Apophis using two-body approximation and full N-body perturbations.}\label{fig_ceu_2bNb}
\end{figure}

Indeed, the main difficulty with the uncertainty parameters is that they are sometimes not indicated. For \textsf{mpcorb}, 105,399 asteroids (that is 17.9 \% of the total in the database) have no U parameter at all and 9431 (1.6 \%) have only a qualitative letter for the U parameter\footnote{In the \textsf{mpcorb} database, the U parameter can be indicated as a letter. If U is indicated as 'E', it means that the orbital eccentricity was assumed. For one-opposition orbits, U can also be 'D' if a double (or multiple) designation is involved or 'F' if an assumed double (or multiple) designation is involved \citep{MPC_U}.}.
In the \textsf{astorb} database, the CEU has not been computed (and appears as 0) for 2585 asteroids (0.44 \%). Moreover, CEU is determined using the approximate two-body problem. For some critical cases (Earth-approaching asteroids), the uncertainties may have been misestimated by a factor of several \citep{astorb}. As an illustration, we have computed the orbital uncertainty - with methods described in the next section - for the asteroid Apophis which is a well known Earth approaching PHA. The difference between the simplified two-body approximation and the full N-body perturbations is clear (see Fig.~\ref{fig_ceu_2bNb}). The ratio between the two values of uncertainty is close to 1 until 2029 and reaches approximately $10^5$ after the 2029-close approach. The 2-body approach does not involve the stretching of the ortbital uncertainty and thus yield to over-optimistic results.

\subsection{Study of the ephemeris uncertainty parameters}
Another problem is that some asteroids have CEU and U parameter in total contradiction. For example, they can have a good CEU (less than 1 arcsec) and a bad U parameter (U=9) (and conversly). To clarify this situation, we have compared the CEU value to the standard deviation $\sigma$ obtained by orbital clones for eight selected representative asteroids (2 with a bad U and bad CEU, 2 with bad U and good CEU, 2 with good U and bad CEU and 2 with good U and good CEU).

In this context, we have computed the standard deviation provided by clones with two different methods using non linear extrapolation. The main principle of these methods is to perform a Monte Carlo propagation of the orbit, \emph{i.e.} to compute clones of initial conditions around the nominal orbit, providing as many orbits as possible. The first method (Monte Carlo using Covariance Matrix, MCCM) consists in adding a random noise to the set of nominal initial conditions using the standard deviation and correlation of these parameters thanks to the covariance matrix. The second one (Bootstrap Resampling, BR) uses the bootstrap resampling method directly on the observations, and consists in determining new set of initial conditions by fitting to a bootstrap set of observations. These two methods, one parametric and one non parametric, have been used previously in \cite{Desmars09} for the study of the precision of planetary satellites ephemerides.

For this test, we have computed 1000 clones for orbital initial conditions of each representative asteroid. Then we have calculated the standard deviation $\sigma$ of the angular separation $s$ in the plane-of-sky at the date of CEU, which represents the angular deviation of an orbit to the nominal one, defined as:
\begin{equation}\label{E:stdv}
\left\{
\begin{array}{rl}
& s_{i}    =  \sqrt{ ((\alpha_{i} - \alpha_{0})\cos \delta_{i})^2 + (\delta_{i} - \delta_{0})^2}\\
&  \sigma  =  \sqrt{(\frac{1}{N-1} \sum_{i}^N s_i^2) - \bar{s}^2}
\end{array}
\right.
\end{equation}

where $i$ denotes the number of the clone, the subscript 0 refers to the nominal orbit, $\bar s$ is the mean, and $N$ is the total number of clones computed.\\

Table \ref{Table_Ceu_clone} gives the comparison between the U parameter, CEU and standard deviation provided by these orbital clones for the two methods.

\begin{table}[h!]
\begin{center}
 \caption{Comparison of the U parameter, the CEU and standard deviation $\sigma$ obtained from orbital clones around the nominal orbits, with two different methods (MCCM and BR, see text). Units are in arcseconds for CEU and $\sigma$, U is unitless.}
\label{Table_Ceu_clone}
\begin{tabular}{ccrrr}
\hline
\hline
   & U	&  CEU  & $\sigma_{MCCM}$ & $\sigma_{BR}$   \\
   & (mpcorb) & (astorb) & (our work)  & (our work)  \\
\hline
 \textbf{2002GM5}     & 9 &  180,000 & 161,461 & 122,517 \\
 \textbf{2006LA}      & 9 &  340,000 & 211,970 & 235,317 \\
\hline	
 \textbf{2000RH60}    & 5 &      0.17 &    0.216 &    0.275 \\
 \textbf{2010JK1}     & 6 &      0.68 &    0.451 &    0.367 \\
\hline
 \textbf{2007WD5}     & 0 &      4300 &  67,388 &  14,844 \\
 \textbf{2003TO9}     & 0 &       240 &      161 &      128 \\
\hline
 \textbf{4321 Zero}   & 0 &     0.058 &    0.077 &    0.086 \\
 \textbf{31824 Elatus}& 0 &     0.36  &    0.548 &    0.331 \\
\hline
\end{tabular}
\end{center}
\end{table}

For these representative asteroids, the CEU is often close to the standard deviation computed here, whereas the U parameter is misestimated for at least four representative asteroids. Moreover, we stress that the U parameter provides a number which is not related to a physical value (a distance or an angle).

In light of the previous tests for several particular cases, and of Fig.\ref{fig_ceu_ceup_u} for the general purpose, the CEU seems to be a good and practical parameter to estimate the accuracy and predictability of an asteroid orbit, as CEU is quickly computable, precise and providing a physical value (an angle).

\section{Relations between CEU and physical or orbital parameters}
\label{S:sec_ARC_CEU}
As the CEU seems to be a useful parameter to measure the orbital uncertainty, we will hereafter use this parameter and highlight several relations between CEU and orbital parameters. 

\subsection{Relation between absolute magnitude, orbital arc and CEU}
Generally, asteroids can be classified in dynamical groups. We propose, for sake of simplicity, to gather the asteroids into the following three main groups defined as\footnote{Number of objects are given at the date of October 5th 2012.}:  

\begin{itemize}
\item \textbf{A-NEA}: asteroid with perihelion $q\le 1.3$ au. It represents asteroids in inner zone of solar system. This group includes the Aten, Apollo, Amor, and Inner Earth Orbit (IEO). There are 9144 objects in A-NEA group in \textsf{astorb}. In this category, Potentially Hazardous Asteroids (PHA) are asteroids presenting a risk of devastating collision (i.e. with $H\le22$ and a Minimum Orbital Intersection Distance, MOID$<0.05$ au).
\item \textbf{B-MBA}: asteroid with perihelion $q > 1.3$ au and semi-major axis $a\le 5.5$ au, representing asteroids in intermediate zone. Thus Trojans are considered as B-MBA in the following. There are 579,208 B-MBAs in \textsf{astorb}.
\item \textbf{C-TNO}: asteroid with perihelion $q>1.3$ au and semi-major axis $a> 5.5$ au, representing asteroids in outer zone of solar system. Thus Centaurs, asteroids between Jupiter and Neptune, are included in C-TNO. There are 1743 C-TNOs in \textsf{astorb}.
\end{itemize}

\begin{figure}[h!]  
\centering 
\includegraphics[width=\columnwidth]{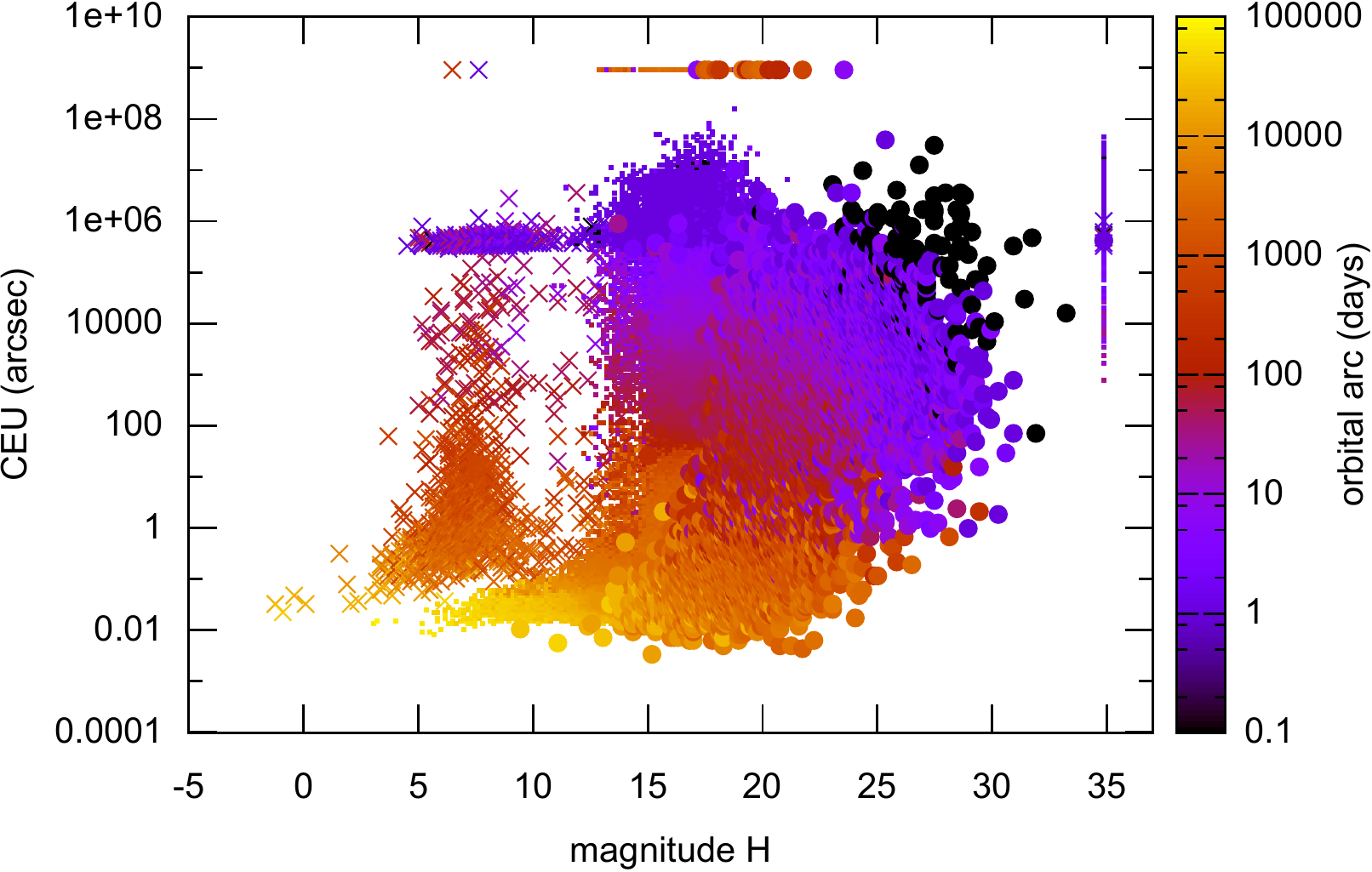}
\caption{{$H$ magnitude, CEU and orbital arc for asteroid from \textsf{astorb}. Cross is for C-TNO, small square for B-MBA and bullet for A-NEA. When $H$ or CEU are unknown, they are represented on the plot with the values $H=35$ or CEU$=10^9$, in contrast to the convention adopted in \textsf{astorb} database where $H=14$.}}\label{fig_H_CEU_ARC}
\end{figure}

Figure \ref{fig_H_CEU_ARC} represents the absolute magnitude $H$, CEU and the length of orbital arc given in \textsf{astorb} for all the asteroids. Each group defined previously appears to be clearly distinguishable: C-TNOs in the left part (cross), B-MBAs in middle part (small square) and A-NEAs in right part (bullet). Large C-TNOs (larger than $\approx$45 km corresponding to absolute magnitude 8, depending on albedo) with short arcs and small A-NEAs (approximately smaller than 5 km whatever the albedo of the asteroid larger than 0.05, using the relation between magnitude and size \citep{MPC_mag_size}) with short arcs also appear. \\

\begin{figure}[h!]  
\centering 
\includegraphics[width=\columnwidth]{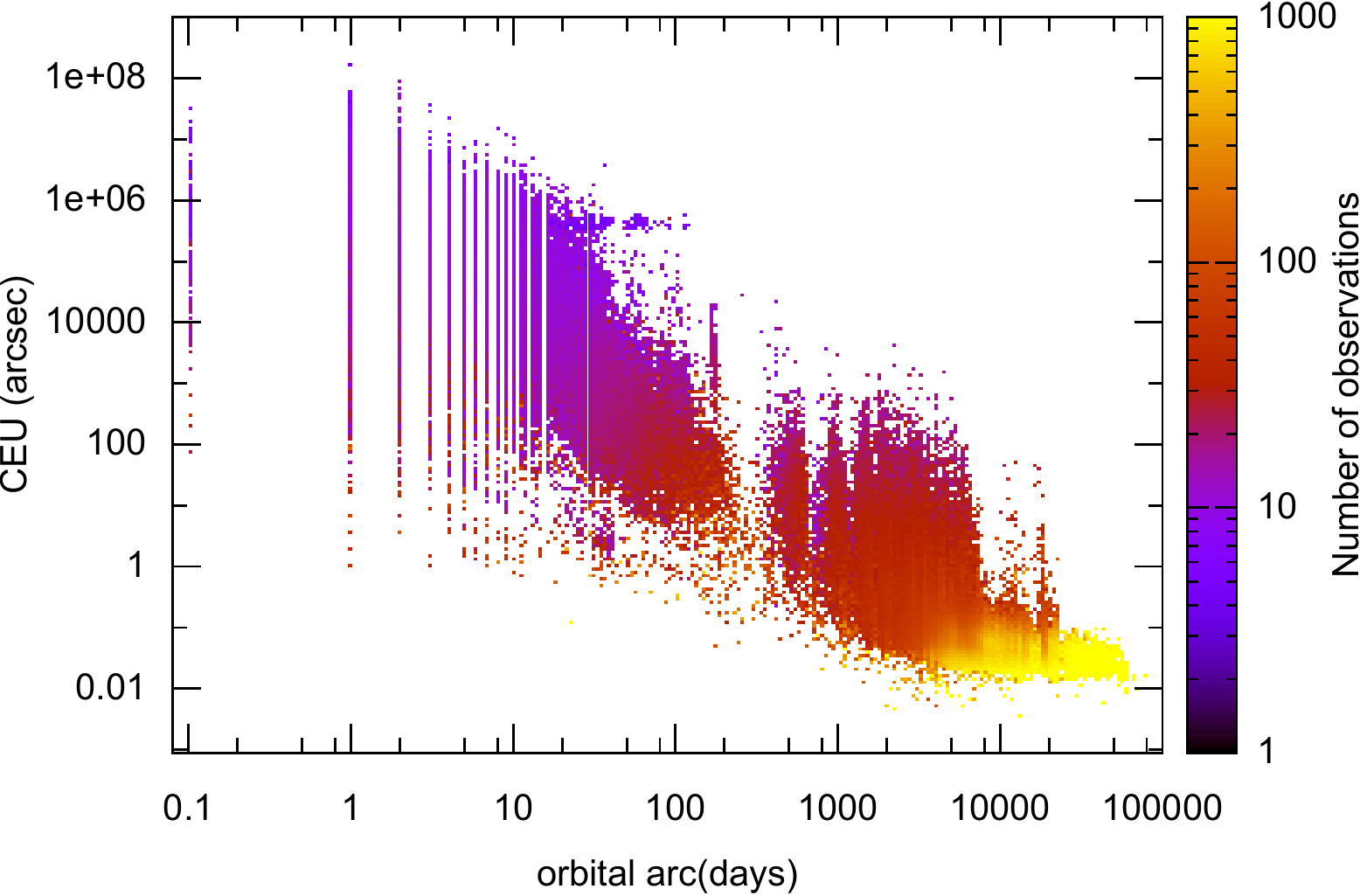}
\caption{Relation between the length of the orbital arc, CEU and the number of observations for all asteroids.}\label{fig_ARC_CEU_NOBS}
\end{figure}

One also notes, that for all the objects, a correlation between the value of CEU, the length of orbital arc and the number of observations can be identified. In Fig.\ref{fig_ARC_CEU_NOBS}, this correlation appears clearly. We can identify three groups related to the length of the arc: 
\begin{itemize}
\item \textbf{arc $<10$ days}: these asteroids are barely observed (less than 10 observations), corresponding mostly to the discovery period, and have a high CEU (between 1 and $10^7$ arcsec). Moreover, the CEU does not decrease much on such graph with increasing orbital arc (see Sect.\ref{Ss:CEUevol} for a discussion of this statement);
\item \textbf{$10$ days $\le$ arc $<250$ days}: these asteroids are not much observed (less than 100 observations) and have a high CEU (between 1 and $10^6$ arcsec);
\item \textbf{$250$ days $\le$ arc $<8000$ days}: these asteroids are also not much observed but have a better CEU (between 0.1 and $100$ arcsec);
\item \textbf{$8000$ days $\le$ arc}: these asteroids have all a large number of observations (more than 100 observations) and have a good CEU (between 0.01 and $10$ arcsec).
\end{itemize}

The gap in orbital arc apparent around 250 days means that when an asteroid is discovered, it is rarely observed for more than 250 days. Beyond this period, the asteroid is in unfavorable configuration to be observed because of its small solar elongation (conjunction) and can be observed again only after its passage behind the Sun. Two peaks can also be observed in Fig.\ref{fig_ARC_CEU_NOBS} at orbital arc longer than 10,000 days. One for arcs of 19,000 days and another, more spread out, for approximately 14,000 days. These peaks can be explained by an important number of objects with such an orbital arc. Actually, many asteroids ($\sim1500$) have been discovered in 1960 with the 1.22m telescope of Mt.Palomar, similarly in 1973. Because these asteroids have been observed recently (in the late 2000s), their orbital arc correspond approximatively to 14,000 days (i.e. the period 1973-2012) and to 19,000 days (i.e. the period 1960-2012). These objects have now various CEU corresponding to the peaks. 

With these considerations, asteroids can be classified in 12 groups considering their orbital arc, their dynamical classes (A-NEA, B-MBA, C-TNO) and their CEU (given the correlation between CEU and length of arc). Table \ref{tab_class} gives the number and the median\footnote{The median is to be prefered to the mean that can be - and generally is - dominated by few large values.} CEU for each group. Asteroids with undetermined CEU (about 2585 asteroids with CEU=0 in \textsf{astorb}) are not taken into account.

\begin{table*}[htdp]
\caption{Different classes of ephemerides precision as given by their median current ephemeris uncertainty (CEU) and length of arc. Numbers of each category are in subscript. Asteroids with undetermined CEU are not taken into account here.}
\label{tab_class}
\begin{center}
\begin{tabular}{rrrr}
\hline
\hline
 & B-MBA$_{[576,647]}$ & A-NEA$_{[9123]}$ & C-TNO$_{[1711]}$ \\
\hline
           arc$<10_{[50,908]}$ & 380,000$_{[48,326]}$ &  6300$_{[2233]}$ & 420,000$_{[349]}$  \\
   $10\le$arc$<250_{[81,540]}$ &      1100$_{[77,579]}$ &   530$_{[3587]}$ & 370,000$_{[374]}$ \\
$250\le$arc$<8000_{[418,331]}$ &     0.11$_{[414,410]}$&  0.16$_{[2951]}$ &      1.0$_{[970]}$ \\
   $8000 \le$ arc$_{[36,731]}$ &    0.045$_{[36,332]}$ & 0.044$_{[352]}$ &       0.13$_{[47]}$ \\
\hline
\hline
\end{tabular}
\end{center}
\end{table*}

\subsection{Relation between orbital arc and CEU}\label{Ss:CEUarc}
\begin{figure}[h!]  
\centering 
\includegraphics[width=\columnwidth]{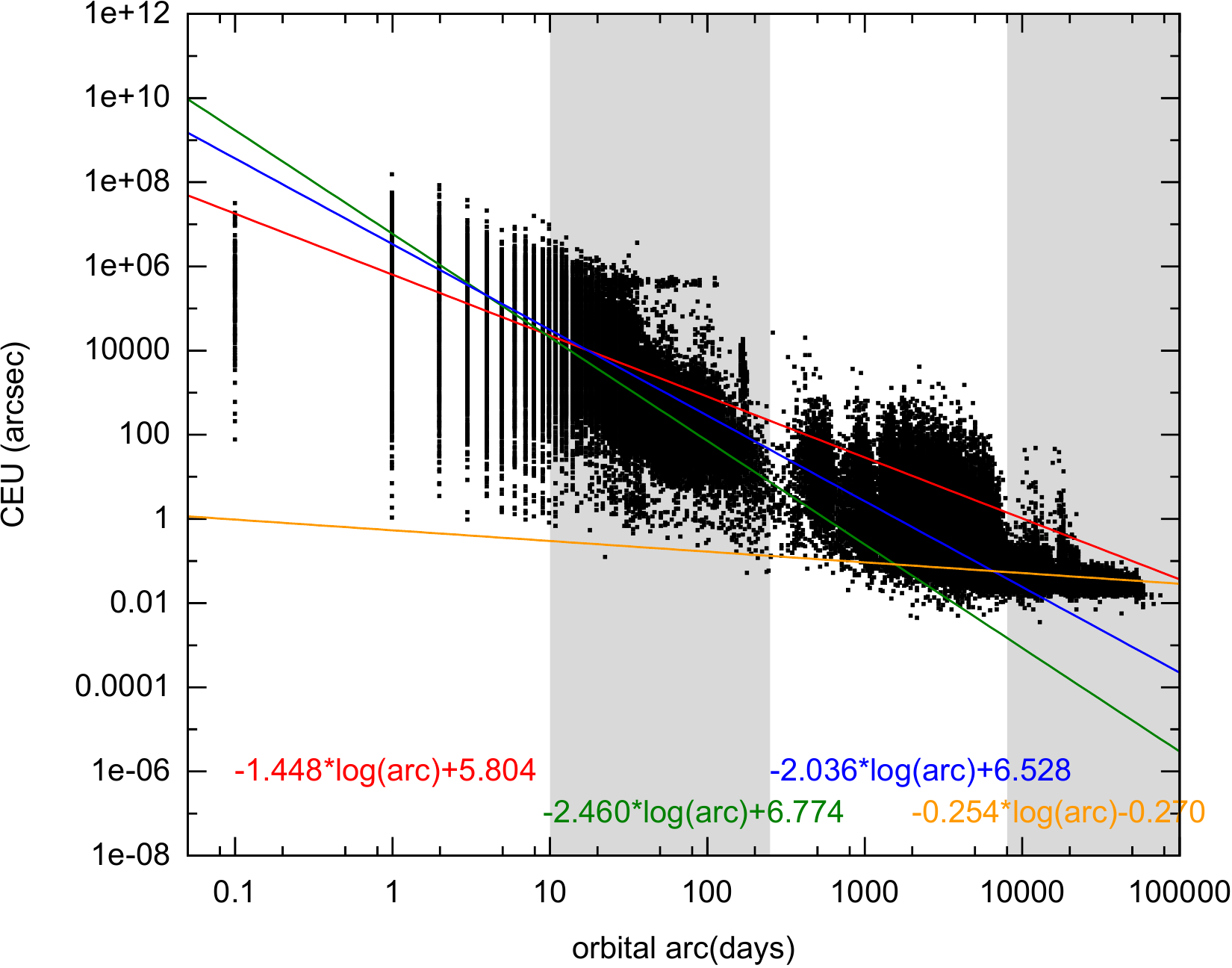}
\caption{Relation between the length of the orbital arc and CEU for all asteroids, and slope for the 4 classes identified (red for arc$<10$d, green for $10\mathrm{d}\le\mathrm{arc}<250$d, blue for $250\mathrm{d}\le\mathrm{arc}<8000$d and orange for $\mathrm{arc}\ge8000$d).}\label{fig_ARC_CEU_reg}
\end{figure}

Figure \ref{fig_ARC_CEU_NOBS} shows that the decrease of the CEU is correlated to the increase of the orbital arc. Moreover the increasing rate of the CEU appears different with varying orbital arc. We have computed an empirical relation for this improvement rate for each of the groups defined in the previous section. For each group, a linear regression can be computed between CEU and the length of orbital arc (as shown in Fig.\ref{fig_ARC_CEU_reg}) and we have the relation:

\begin{equation}
\log(CEU) = a\log(arc)+b
\end{equation}
where $a$ and $b$ are given in Table \ref{tab_ARC_CEU_reg} and Fig.\ref{fig_ARC_CEU_reg} for each groups. \\

If the orbital arc is smaller than 10 days\footnote{Orbital arcs indicated as 0 in the database are considered as 0.1 in this section to allow the use of logarithmic scale.}, the CEU is very large and does not improve much when the length of arc becomes larger (see also Sect.\ref{Ss:CEUevol}). However, in such a case, observations remain useful to avoid loosing the object. Between 10 and 250 day arcs, the CEU is clearly improved when the length of arc increases. For asteroids with an orbital arc between 250 and 8000 days, the CEU is smaller than 100 arcsec and is still much improved when the length of arc becomes greater. If the orbital arc is greater than 8000 days, then the CEU is not much improved and reaches its typical minimum value of about 0.1-0.2 arcsec.\\ 

\begin{table}[h!]
\begin{center}
\caption{Values of the linear regression $\log(CEU) = a\log(arc)+b$, for each groups defined in Table\ref{tab_class}.}
\label{tab_ARC_CEU_reg}
\begin{tabular}{rrrr}
\hline
\hline
\textbf{arc} (days) & $a$ & $b$ & \textbf{number}\\
\hline
0.1-10		& -1.448  & +5.804 &  50,908 \\
10-250		& -2.460  & +6.774 &  81,540 \\
250-8000 	& -2.036  & +6.528 & 418,331 \\
$>8000$	& -0.254  & -0.270 &  36,731 \\
\hline
\end{tabular}
\end{center}
\end{table}

As noted previously, in the short arc group (orbital arc less than 10 days) the CEU does not decrease quickly while the orbital arc increases. It is unclear however if this is associated to the population on average or on each individual arc. In the latter case it could be a surprising result; so we highlight the trend on specific study case in the following section.
 
\subsection{Evolution of CEU for short orbital arc}\label{Ss:CEUevol}
In this section, we investigate the evolution of the CEU as a function of the orbital arc, for a single object. We try to highlight the time spent after the discovery to really improve the quality of the orbit. A previous work in \cite{virtanen06} studied the time evolution of the orbital elements uncertainty of the asteroid 2004 AS1 together with the evolution of the impact risk assessment. For this purpose, the authors used the volume of variation (VoV) technique which is a nonlinear Monte Carlo technique \citep{muinonen06}. 

\begin{figure}[h!]  
\centering 
\includegraphics[width=\columnwidth]{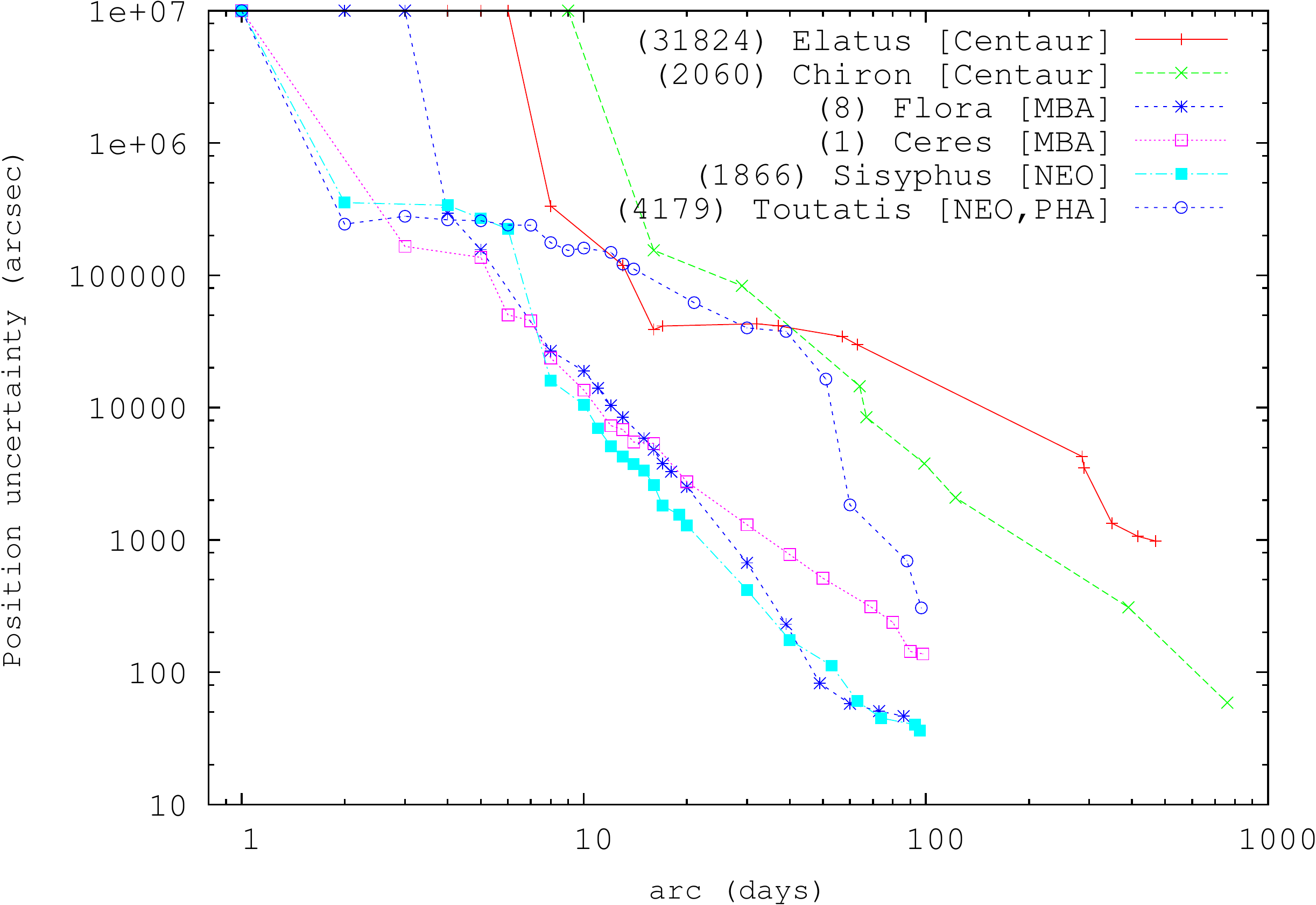}
\caption{Variation of the orbital uncertainty as a function of the arc length for six single objects. The axis are expressed in a logarithmic scale.}\label{F:variation_ceu}
\end{figure}

In this context, we computed the orbital uncertainty of an asteroid by considering a variable (and short) length of orbital arc. For our work, six test case objects were considered: (31824) Elatus and (2060) Chiron for Centaur asteroids; (1) Ceres and (8) Flora for MBAs; (1866) Sisyphus and (4179) Toutatis (PHA) for NEOs. First, we performed a preliminary orbit determination using the Gauss method. The Gauss method converged two days after discovery for all cases except for Chiron and Ceres which converged one day after discovery. To assess the uncertainty of the orbit, we used the differential correction method. All of these computations were performed using OrbFit4.2 package\footnote{\url{http://adams.dm.unipi.it/orbfit/}}. Then, for each arc, we computed 1001 clones of the nominal orbit from a multivariate Gaussian distribution given by the covariance matrix (obtained from the procedure described above).
Each clone orbit is further directly propagated at the date of computation of CEU with Monte Carlo technique. The Lie integrator \citep{bancelin12} was used to perform this propagation.
Thus, we were able to compute the standard deviation $\sigma$ using (Eq. \ref{E:stdv}) that we assumed to match with the definition of the CEU. If the Gauss method and/or orbital improvement failed or if the $\sigma$ value obtained is large (greater that $10^7$), then the CEU is considered as indeterminate and is set to $10^7$. Obviously, some asteroids considered in this study have an indeterminate CEU at the epoch of their discovery (Fig. \ref{F:variation_ceu}). For Chiron and Ceres, a preliminary solution was obtained but the uncertainty on the orbit was too large. For the NEOs, the Gauss method converged 2 days after the discovery and the differential correction succeeded. For the MBAs, it is defined after 3 days whereas for the Centaurs, the uncertainty is not defined up to ten days. Then, the orbital uncertainty drops almost linearly for all the objects: for the Centaur asteroids, it goes below 100-1000 arcsec for an arc length 800-1000 days. This is due to the lack of observations because they are distant objects and they have a slow apparent motion. For the other asteroids, the orbital uncertainty decreases linearly and goes below 100 arcsec. 

One can see that its variation for the asteroid Toutatis has a stranger behaviour than the others because the sharp drop of the uncertainty value occurs around the 14th day. 

This test shows that, except for the distant objects (Centaurs), observations performed after their discovery (until $\sim$ 6 days) will mostly avoid the loss of the object than improve its orbit, as the rate of change is not that significant. However, continuous observations performed after $\sim$ 6 days will allow a significant orbit improvement as the rate is much steeper.

\section{Astrometric measurements}\label{S:measurements}
In this section, we first described the different types of measurement of asteroid position, then we estimate the effect of specific measurements (optical observations, radar or future Gaia observations) in the improvement of the orbit. 
\subsection{Presentation}
The position of an asteroid can be determined by many types of measurements. The Minor Planet Center provides all the measurements related to asteroids for numbered and unnumbered asteroids \citep{MPC_CATOBS}. As of September 30th 2012, 94,082,128 observations are available on the database. Table \ref{tab_MPCOBS} gives the statistics about the different kind of measurements provided by the MPC database in particular their percentage and the time span. 

Historically, the first observations of asteroids were made with micrometer during the 19\textsuperscript{th} century. In the latter part of that century, the photographic plates allowed a better measurement of asteroid position. In the mid-1980s, the CCD has revolutionized astrometric measurement thanks to a better sensitivity, stability, dynamic range and ease of extracting astrometric measurements. Consequently, most of the measurement in the MPC database come from CCD. Space observations such as by Hubble Space Telescope, Spitzer and particularly WISE also provide an important number of measurements. 

In addition to the different types of measurements mentioned above, some observations can appear as secondary or marginal for different reasons; even if being precise they can have a lower impact on the orbital models. For example, the space mission Hipparcos also provides geocentric data \citep{Hestroffer1998}. Radar measurements are not very common but give the most accurate measurements for asteroid. They also can provide physical characteristics such as shape. Astrometric positions can also be deduced from stellar occultations.\\

\begin{table*}[htdp]
\begin{center}
\caption{Statistics of different kind of measurement provided by MPC database for numbered and unnumbered asteroids. }\label{tab_MPCOBS}
\begin{tabular}{cp{4cm}lrrc}
\hline
\hline
\textbf{code} & \multicolumn{2}{l}{\textbf{type}} &  \textbf{number} & \textbf{percentage} &  \textbf{timespan} \\
\hline
\rowcolor{gris}  C &  \multicolumn{2}{> {\columncolor{gris}}l}{CCD}         &          88,546,921 &    94.12\% &   1986-2012 \\ 
                S/s & \multicolumn{2}{l}{Space observation}                &           4,006,572 &     4.26\% &   1994-2011 \\ 

                    &  & \scriptsize\emph{HST}            &         \scriptsize\emph{3544} &     \scriptsize\emph{0.00\%} &   \scriptsize\emph{1994-2010} \\ 
                    &  & \scriptsize\emph{Spitzer}        &          \scriptsize\emph{114} &     \scriptsize\emph{0.00\%} &   \scriptsize\emph{2004-2004} \\ 
                    &  & \scriptsize\emph{WISE}           &  \scriptsize\emph{4,002,914} &     \scriptsize\emph{4.25\%} &   \scriptsize\emph{2010-2011} \\ 

\rowcolor{gris}  A & \multicolumn{2}{> {\columncolor{gris}}l}{Observations from B1950.0 converted to J2000.0}  &          647,690 &     0.69\% &   1802-1999 \\ 
                  c & \multicolumn{2}{l}{Corrected without republication CCD observation}                         &          462,065 &     0.49\% &   1991-2007 \\ 
\rowcolor{gris}  P & \multicolumn{2}{> {\columncolor{gris}}l}{Photographic}                                    &          352,449 &     0.37\% &   1898-2012 \\ 
                  T & \multicolumn{2}{l}{Meridian or transit circle}                                              &           26,968 &     0.03\% &   1984-2005 \\ 
\rowcolor{gris}X/x & \multicolumn{2}{> {\columncolor{gris}}l}{Discovery observation}                           &           16,741 &     0.02\% &   1845-2010 \\ 
                  M & \multicolumn{2}{l}{Micrometer}                                                              &           12,081 &     0.01\% &   1845-1954 \\ 
\rowcolor{gris}  H & \multicolumn{2}{> {\columncolor{gris}}l}{Hipparcos geocentric observation}                &            5,494 &     0.01\% &   1989-1993 \\ 
                  R & \multicolumn{2}{l}{Radar observation  }                                                     &            2,901 &     0.00\% &   1968-2012 \\ 
\rowcolor{gris}  E & \multicolumn{2}{> {\columncolor{gris}}l}{Occultations derived observation}                &            1,737 &     0.00\% &   1961-2012 \\ 
                  V & \multicolumn{2}{l}{"Roving observer" observation}                                           &               388 &     0.00\% &   2000-2012 \\ 
\rowcolor{gris}  n & \multicolumn{2}{> {\columncolor{gris}}l}{Mini-normal place derived from averaging observations from video frames}  &               105 &     0.00\% &   2009-2012 \\ 
                  e & \multicolumn{2}{l}{Encoder}                                                                 &                16 &     0.00\% &   1993-1995 \\ 
\hline
\hline
\end{tabular}
\end{center}
\end{table*}

The accuracy of each measurement can be obtained thanks to the AstDyS website \citep{astdys}. For each numbered asteroid, this website provides an observation file indicating the residual (O-C) of each measurement. All the observations of all numbered asteroids\footnote{As from October 5th 2012, there are 337,008 numbered asteroids in AstDyS database.} have been compiled and the accuracy of each measurement has been estimated by computing a weighted\footnote{AstDys provides a weight for each observation according to their accuracy.} root mean square of all residuals of the measurement. 

Table \ref{tab_MPCOBS_accuracy} provides the number, the percentage of accepted measurement and the estimated accuracy for each kind of observations. After radar measurement, observations derived from occultations and Hipparcos geocentric observations appear as the more accurate one (about 0.08-0.15 arcsec).  Recent measurements such as CCD observations or meridian circle observations are rather accurate (between 0.2-0.4 arcsec). Older measurements such as photographic or micrometric measures are less accurate (1.0-1.8 arcsec). Surprisingly, observations from spacecraft (HST, Spitzer or WISE) are not any more accurate than groundbased data. Nevertheless, Wide-field Infrared Survey Explorer (WISE) was designed for asteroid detection and actually, the mission has detected 157,000 asteroids, including more than 500 NEO \citep{Mainzer2011}.

\begin{table*}[htdp]
\begin{center}
\caption{Accuracy of measurements for numbered asteroids from AstDyS2 database.}\label{tab_MPCOBS_accuracy}
\begin{tabular}{clrrrl}
\hline
\hline
\multirow{2}{*}{\textbf{type}} & \multirow{2}{*}{\textbf{name}} & \textbf{number of}  &  \textbf{percentage of} & \multicolumn{2}{c}{\multirow{2}{*}{\textbf{accuracy}}} \\
  &  & \textbf{measurement}  &  \textbf{accepted measurement} & &  \\
\hline
  C & CCD                       &    79,569,190 &   99.49\% &     0.388 & arcsec  \\
  S & Wise                      &     1,526,466 &   99.86\% &     0.583 & arcsec  \\
  S & Hubble Space Telesc.      &           867 &   96.54\% &     0.585 & arcsec  \\
  S & Spitzer                   &            48 &   33.33\% &     1.673 & arcsec  \\
  A & B1950 to J2000            &       632,428 &   82.17\% &     1.170 & arcsec  \\
  c & Corrected CCD obs.        &       423,792 &   99.70\% &     0.507 & arcsec  \\
  P & Photographic              &       346,947 &   93.38\% &     1.084 & arcsec  \\
  T & Meridian/transit circle   &        26,968 &  100.00\% &     0.250 & arcsec  \\
  M & Micrometer                &        12,081 &   94.56\% &     1.742 & arcsec  \\
  X & Discovery obs.            &          9520 &    0.81\% &     0.898 & arcsec  \\
  H & Hipparcos obs.            &          5494 &  100.00\% &     0.148 & arcsec  \\
  E & Occultations              &          1736 &  100.00\% &     0.085 & arcsec  \\
  R & Ranging                   &           612 &   96.41\% &     3.325 & km  \\
  R & Doppler                   &           432 &   99.07\% &     6.660 & km/s  \\
  V & Roving observer           &           372 &   49.73\% &     0.822 & arcsec  \\
  e & Encoder                   &            16 &  100.00\% &     0.558 & arcsec  \\
  n & video frame               &            12 &  100.00\% &     0.319 & arcsec  \\
\hline
\hline
\end{tabular}
\end{center}
\end{table*}

\subsection{Impact of astrometric measurements on orbit uncertainty - The Apophis example}
In order to quantify the impact of a type of measurement on the orbital uncertainty
of an asteroid, we specifically dealt with asteroid (99942) Apophis. This particular object belongs to the PHAs family and is known to be the most
threatening object of the last decade since it is the only asteroid who reached level 4 of Torino scale for a potential impact with the Earth on
April 2029 \citep{sansaturion08}. Since, new observations (optical and radar) ruled out every possibility of impact in 2029 and this threat turned out
to be a close encounter within 5.64R$_{\oplus}$ with Earth. But this close encounter is so deep that the asteroid will move from the Aten family to
the Apollo family orbit. Besides, its orbit will become chaotic and  new possibilities of collision with the Earth after 2029 will appear. The most consensual
collision date is in 2036 for which, at the date of the last observations available in 2012, the risk was estimated with a probability of $\sim
10^{-6}$. This date is also famous because of the size of \textit{keyhole}, region in the B-plane \citep{Valsecchi2003} of 2029 where the asteroid has to
pass in order to collide the Earth in the future. The 2036-keyhole size is estimated at $\sim 600$m. In the B-plane, we cannot only represent the
state of the asteroid, located by two geocentric coordinates ($\xi$, $\zeta$), at the date of close encounter, but also its uncertainty
($\sigma_{\xi}$, $\sigma_{\zeta}$) and the relative position of the keyhole with respect to the nominal solution. More often than not, the uncertainty
region is a 3$\sigma$-ellipse centered on the nominal solution and its size is directly linked to the observations used. Therefore, the position of
keyhole compared with the size of the ellipse uncertainty is also important to quantify the risk of future collision \citep{chesley06,kochetova09,bancelin12a}.\\

\subsubsection{Past observations}

\begin{figure}
\centering
 \includegraphics[width=\columnwidth]{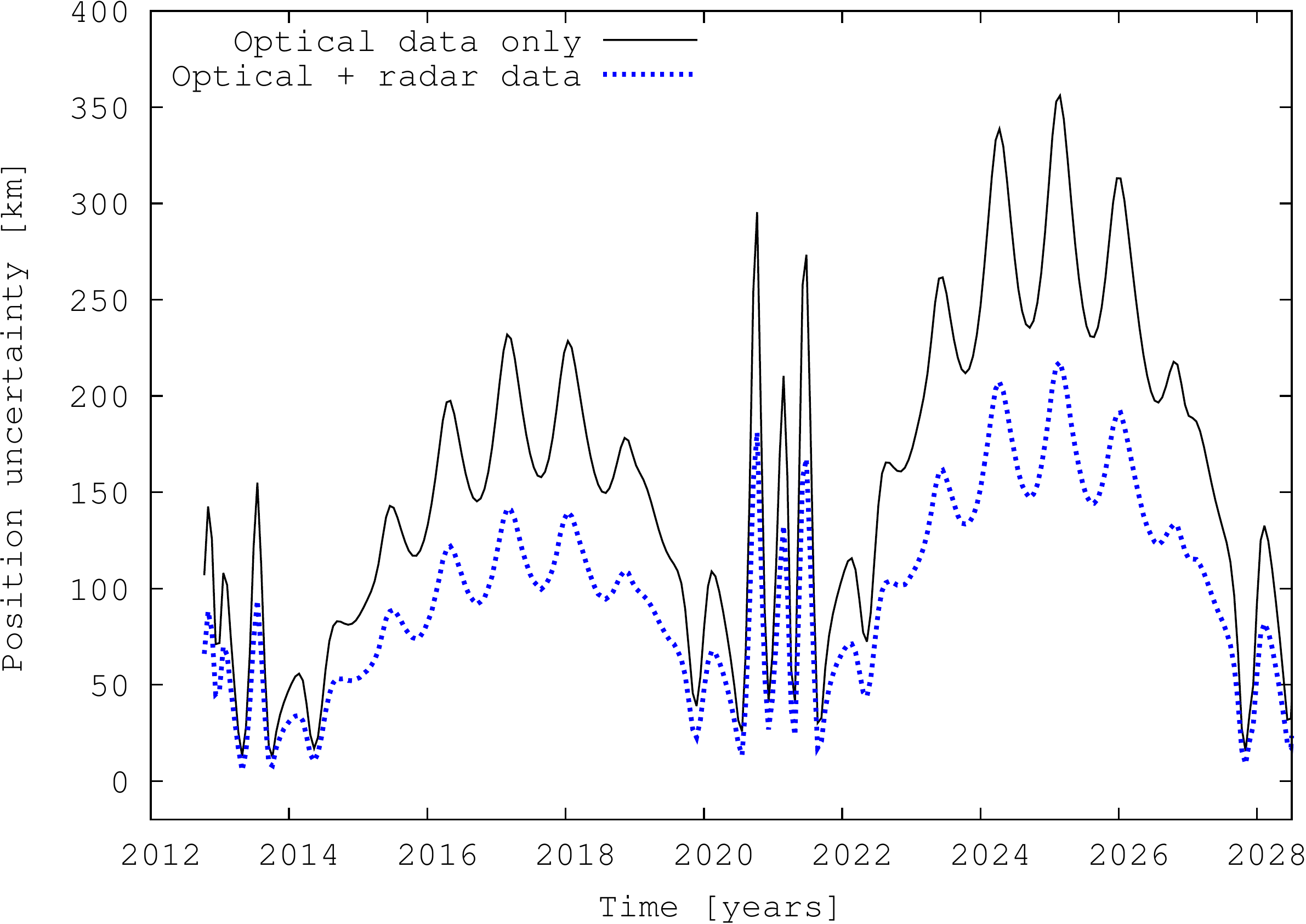}
\caption{Position uncertainty on the geocentric distance of Apophis after fitting to optical observations (solid line) and optical and radar data
observations (dashed line).}\label{F:current_uncer}
\end{figure}

In this section, we first deal with the impact of the observations of Apophis (from 2004-2008), as available at MPC-UAI, on the
uncertainty region. We propagated the equations of motion (as well as the variational equations) of Apophis until 2029, considering gravitational
perturbations of all the planets, including also the gravitational perturbations  of the Moon, Ceres, Pallas and Vesta. Relativistic accelerations
were also taken into account. The uncertainty of the nominal orbit was propagated using a linear propagation of the initial covariance matrix in
order to assess the evolution of the orbit uncertainty. We performed this test using two different sets of observations. The first
one uses only optical data from 2004-2008 and the second one uses optical and radar data (5 radar measurements -- ranging and Doppler measurements --
were performed at Arecibo in 2005 and 2006). This test is done in order to highlight the impact of radar data on the current orbit uncertainty.\\
First, we represent in Fig. \ref{F:current_uncer}, the current position uncertainty of Apophis until the date of close encounter, when considering
optical data only or optical and radar data. One can see that the uncertainty is improved by a factor of more than 2 when considering radar data. This
is not a surprise in as much as, radar data are very accurate measurements. 

This improvement can also be illustrated by the evolution of the uncertainty on the 2029 B-plane.
Table \ref{T:b_plane_uncertainty2} shows the value of the nominal position ($\xi$, $\zeta$) and the uncertainty ($\sigma_{\xi}$, $\sigma_{\zeta}$).
The uncertainty on the geocentric distance $\Delta$ is also given through the value of $\sigma_{\scriptscriptstyle \Delta}$. One can see that radar
measurements enable to decrease the uncertainty on the main component $\zeta$ and on $\Delta$ by a factor more than 2, while the uncertainty on $\xi$ component remains almost unchanged.

\begin{table}[h!]
\begin{center}
 \caption{Uncertainty ($\sigma_{\scriptscriptstyle \xi}$,$\sigma_{\scriptscriptstyle \zeta}$) on the 2029 B-plane of Apophis considering optical data
only or optical and radar data.}
 \label{T:b_plane_uncertainty2}
 \begin{tabular}{ccc}
  \hline
  \hline
           & Optical + radar data & Optical data only  \cr
 \hline
  $ \xi \pm \sigma_{\scriptscriptstyle \xi}$ (km)& 6980 $\pm$ 15& 6963 $\pm$ 17\cr
 \hline
  $ \zeta \pm \sigma_{\scriptscriptstyle \zeta}$ (km) & 37,440 $\pm$ 345& 37,144 $\pm$ 788\cr
 \hline
  $ \Delta \pm \sigma_{\scriptscriptstyle \Delta}$ (km) & 38,068 $\pm$ 345& 37,791 $\pm$ 788\cr
  
\hline
\hline
\hline
 \end{tabular}
\end{center}
\end{table}

One should note that if the arc data increases thanks to optical measurements and without radar data, the orbital improvement using ranging and/or doppler measurements could not be that significant. For instance, Apophis has additional optical observations from 2011-2012. Considering those data combined with the radar measurements, the uncertainty on the distance drops to 154 km. If we ignore the radar data, this uncertainty value is 177 km.

\subsubsection{Future observations}

Gaia is an astrometric mission that will be launched in autumn 2013. The aim is to have a 3D-map of our Galaxy. There are many scientific outcomes
 from this mission and as far as our Solar System is concerned, the satellite will be able to map
thousands of main belt asteroids (MBAs) and near-Earth objects (NEOs) down to magnitude 20. The
high precision astrometry (0.3-5 mas of accuracy on a transit level) will allow orbital improvement, mass determination,
 and a better accuracy in the prediction and ephemerides of potentially hazardous asteroids (PHAs). During the 5-year mission, Gaia may observe
several time the asteroid Apophis \citep{Bancelin2012}. Thereafter, we performed some tests to quantify the impact of future measurements (optical, radar or space data) on
Apophis orbit when the new data
are outside the period of observations. Apophis will have some favorable conditions of observations in 2013 and 2021 as it will be at a distance
of less than 0.1 au In this context, we have considered 5 sets of possible measurements:
\begin{itemize}
\item $S_1$: using all optical and radar data available (2004-2012);
\item $S_2$: using set $S_1$ with additional Gaia data (over the period 2013-2018) with 5 mas accuracy;
\item $S_3$: using set $S_1$ with one additional future radar measurement in 2013 with 1$\mu$s accuracy (measurement of a timing echo);
\item $S_4$: using set $S_1$ with one future optical observation done in 2013 with 0.1 arcsec accuracy;
\item $S_5$: using set $S_1$ and the case that Gaia would provide only one observation with 5 mas accuracy.
\end{itemize}

\begin{figure}[h!] 
\centering 
\includegraphics[width=\columnwidth]{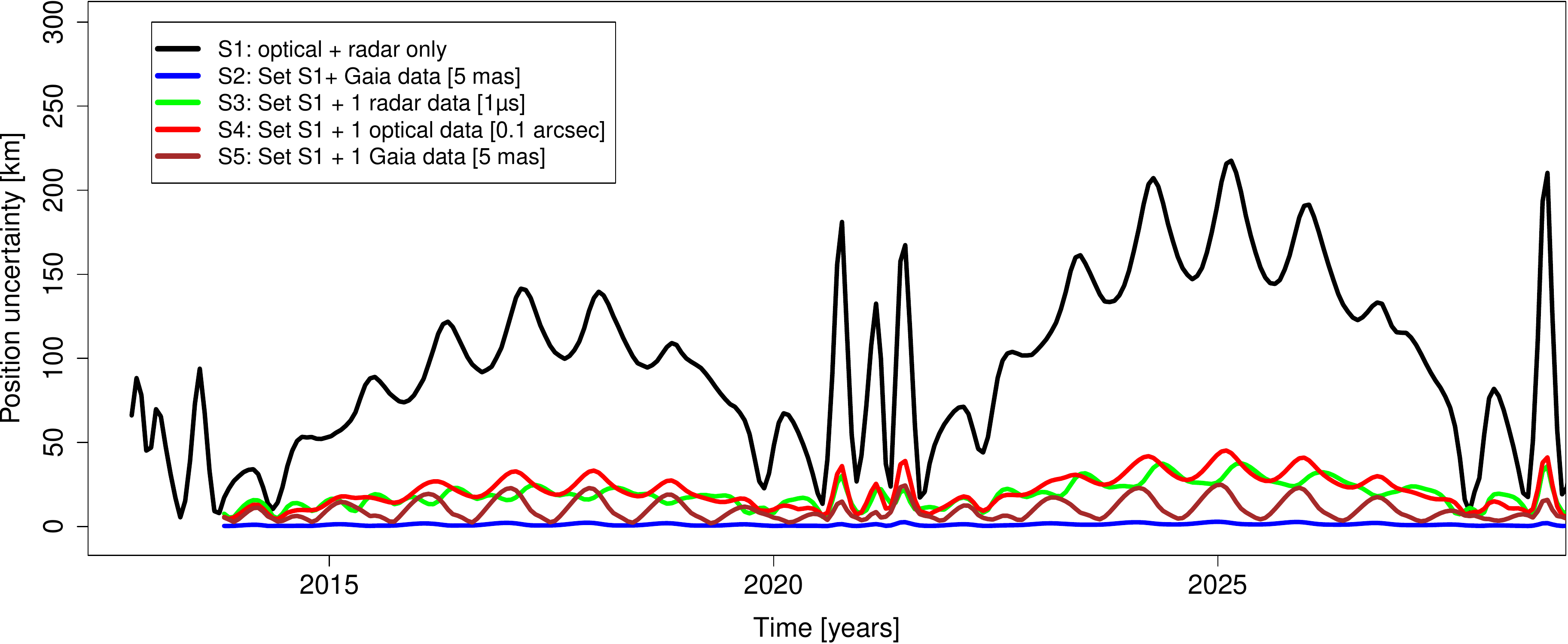}
\caption{Evolution of the position uncertainty (given by its geocentric distance) of asteroid Apophis considering several different sets of
observations.}\label{F:uncertainty} 
\end{figure}

Figure \ref{F:uncertainty} shows that the Gaia data enable to reduce the position uncertainty knowledge to the kilometer level (set $S_2$)
and to keep this value until the close approach of 2029. For comparison, the effect of future accurate measurements (radar and optical) can be
comparable to the
impact of one future Gaia data. 

In the same context, we have quantified the uncertainty in the B-plane of 2029. While the impact of one additional Gaia data (set $S_5$) can be
compared to
the effect of one radar measurement (set $S_3$), one set of Gaia observations (set $S_2$) can bring the
uncertainties around the kilometer level (Tab.\ref{T:b_plane_uncertainty}).
\begin{table}[h!]
\begin{center}
 \caption{Uncertainty ($\sigma_{\scriptscriptstyle \xi}$,$\sigma_{\scriptscriptstyle \zeta}$) on the 2029 B-plane of Apophis considering various sets
S$_{\scriptscriptstyle i}$ of future observations (see text). }
 \label{T:b_plane_uncertainty}
 \begin{tabular}{|c|c|c|c|c|c|}
  \hline
  \hline
           & S$_{\scriptscriptstyle 1}$ & S$_{\scriptscriptstyle 2}$ & S$_{\scriptscriptstyle 3}$ & S$_{\scriptscriptstyle 4}$ &
S$_{\scriptscriptstyle 5}$    \cr
 \hline
  $\sigma_{\scriptscriptstyle \xi}$ (km)& $3.45$  & $0.23$ & $3.38$ & $3.44$ & $1$\cr
 \hline
  $\sigma_{\scriptscriptstyle \zeta}$ (km) & $154$ & $2.7$ & $10.93$ & $47$ & $19.8$\cr 
\hline
\hline
 \end{tabular}
\end{center}
\end{table}

\subsection{Impact of Gaia stellar catalogue on orbit uncertainty}
The space mission Gaia will not only provide accurate observations of asteroid positions but also and especially a astrometric stellar catalogue. The Gaia catalogue \citep{mignard07} will provide unbiased positions of a billion of stars until magnitude 20 and with an accuracy depending on magnitude (7 $\mu$as at $\le$10 mag; 12-25 $\mu$as at 15 mag and 100-300 $\mu$as at 20 mag). With this stellar catalogue, a new astrometry era will be possible.
New process of reduction of new and archived observations will be necessary such as propagation model to third order for star proper motions, differential aberration, atmospheric absorption, color of stars, etc. With such corrections, an accuracy of 10 mas is expected on the position of any asteroid deduced from Gaia stellar catalogue. In particular, it will be possible to reduce future but also old observations with this catalogue even if it seems obvious that all the observations for all the asteroids could not be reduced again.\\

In this section, we studied the impact of this new reduction in the orbit uncertainty and we considered the case of (99942) Apophis and assumed its current observations: 1518 optical observations and 7 radar measurements from 2004 to 2012. 

\begin{figure}[h!]  
\centering 
\includegraphics[width=\columnwidth]{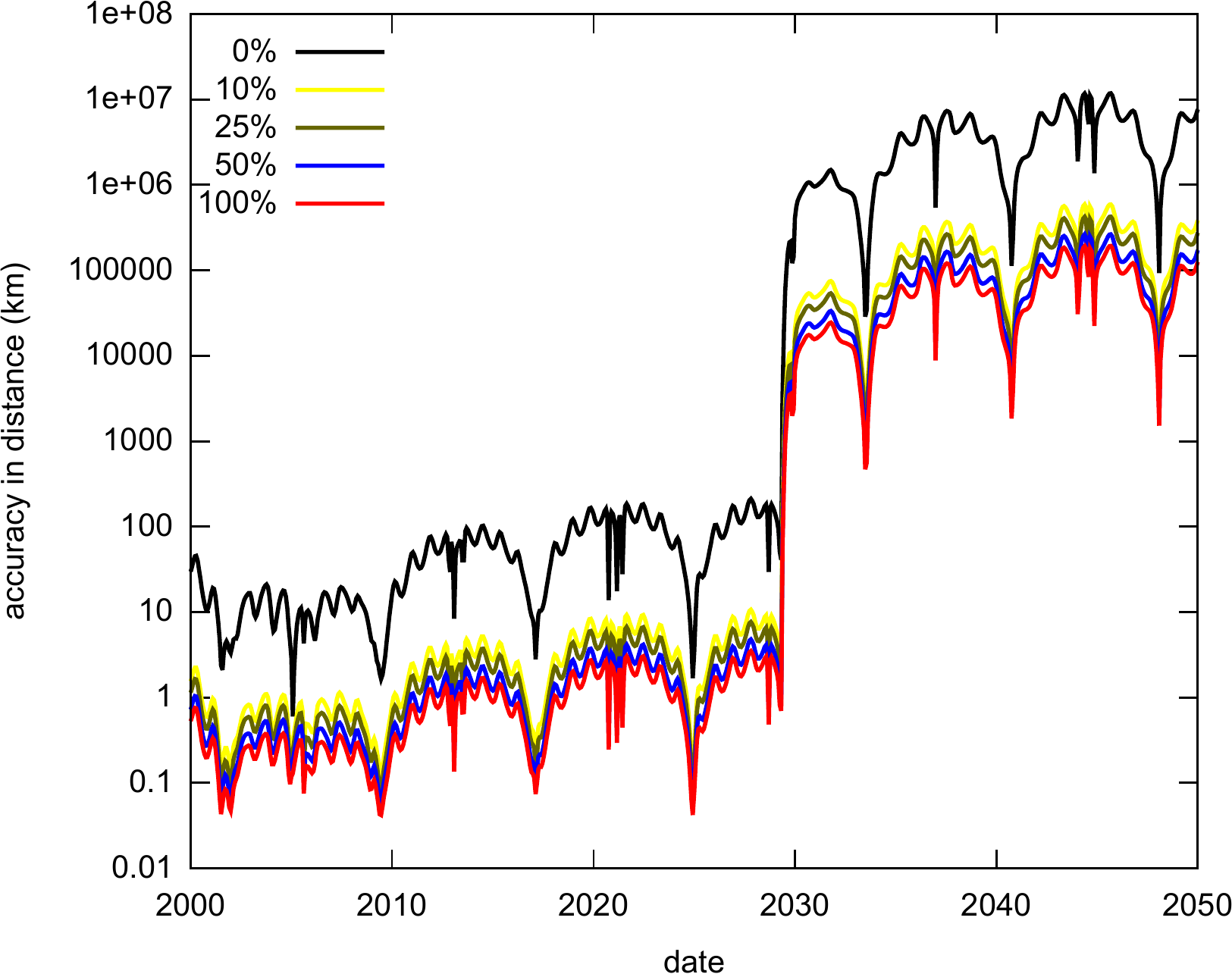}
\caption{Evolution of the position uncertainty of asteroid Apophis by considering that a certain percentage of optical observations could be reduced with Gaia stellar catalogue, i.e. with an accuracy of 10 mas (see text).}\label{fig_Apophis_gaiarandom}
\end{figure}

We assumed that a variable part of the optical observations could be reduced again with the Gaia stellar catalogue (i.e. with an accuracy of 10 mas systematically) and we computed on figure \ref{fig_Apophis_gaiarandom}, the position uncertainty from 2000 to 2050 depending on a percentage of reduced observations with Gaia stellar catalogue (0\%, 10\%, 25\%, 50\% and 100\%). The case of 0\% means that none of the optical observations could be reduced again and represents the current position uncertainty by considering the current accuracy of the observations. 100\% means that all the observations could be reduced with Gaia stellar catalogue.

The variation of the uncertainty is the same for all the cases but the amplitude is different (Fig.~\ref{fig_Apophis_gaiarandom}). With the current accuracy of observations, the uncertainty of the Apophis orbit is about 10 to 100 km in geocentric distance from 2000 to 2029 and then sharply increases after the 2029 Earth close approach. While the percentage of observations reduced with Gaia stellar catalogue increases, the uncertainty of the orbit decreases. Figure \ref{fig_Apophis_gaiarandom} shows that if only 10\% of observations were reduced with Gaia stellar catalogue, the accuracy of the position would be 20 times smaller than with current observations. If the percentage reaches 50\%, the position uncertainty would be 40 times smaller. Finally, if all the optical observations of Apophis could be reduced again with Gaia catalogue then the orbit accuracy would be 50 times better than with current observations. \\

In the case where just a few number of observations can be reduced again, the choice of the first and last observations appears as opportune. Indeed, we have compared this case with a random choice of the same number of observations reduced by Gaia stellar catalogue. The selection of the first and last observations provides the best accuracy. In fact, the first and last observations, by virtually increasing the length of orbital arc, bring more important constraints on the orbital motion of the asteroid.

Consequently, we assumed that a small number of the optical observations could be reduced again with the Gaia stellar catalogue and we computed the accuracy of the orbit from 2000 to 2050. We dealt with five situations:
\begin{itemize}
\item current observations with their current accuracy;
\item we assumed that the first and the last observations can be reduced with Gaia stellar catalogue i.e. with a conservative accuracy of 10 mas;
\item we assumed that the first five and the last five observations can be reduced with Gaia stellar catalogue;
\item we assumed that the first ten and the last ten observations can be reduced with Gaia stellar catalogue;
\item we assumed that all observations can be reduced with Gaia stellar catalogue.
\end{itemize}

\begin{figure}[h!]  
\centering 
\includegraphics[width=\columnwidth]{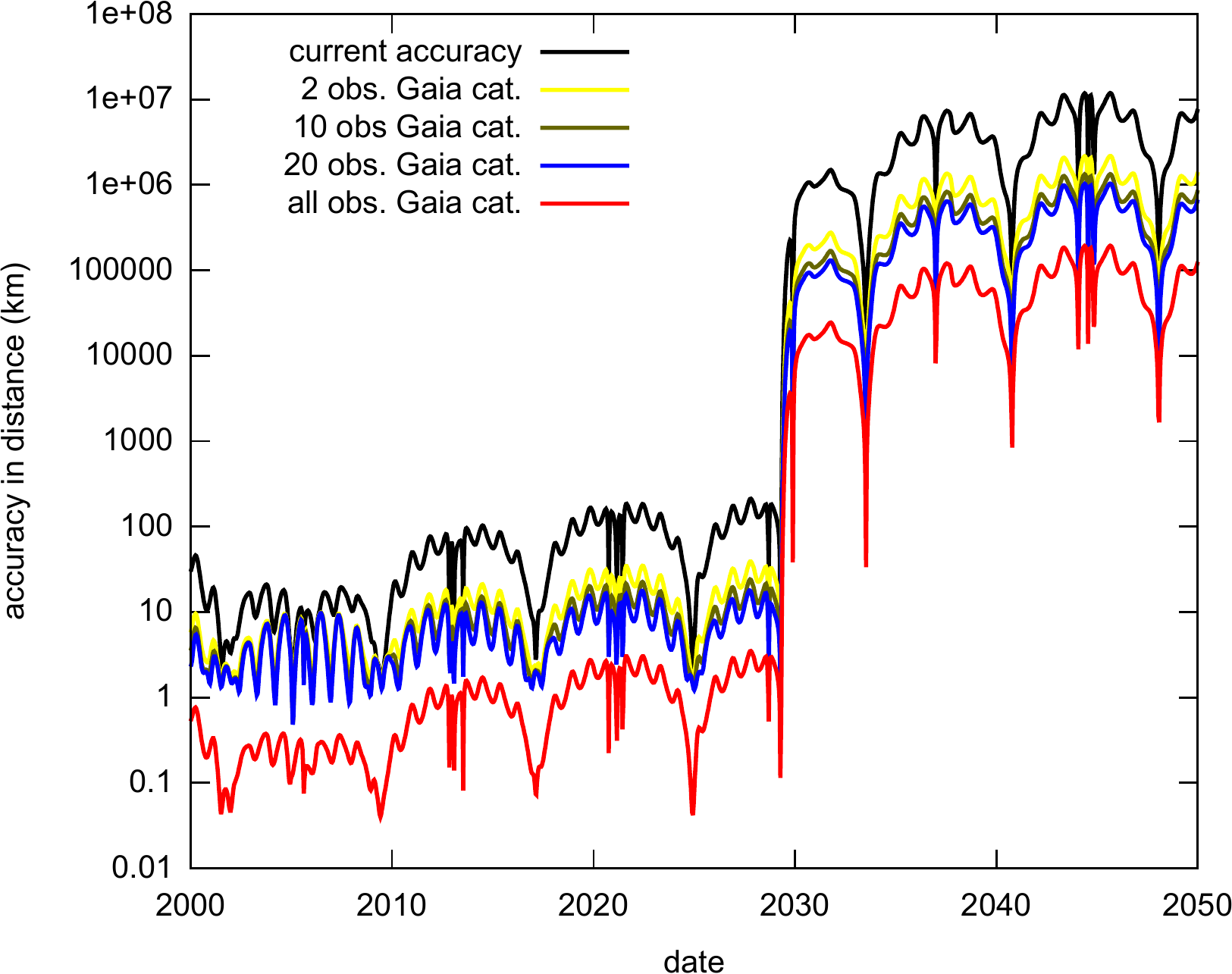}
\caption{Evolution of the position uncertainty of asteroid Apophis considering that a few number of first and last observations could be reduced by Gaia stellar catalogue (see text).}\label{fig_Apophis_gaiacat}
\end{figure}

Figure \ref{fig_Apophis_gaiacat} shows that even if only a few number of first and last observations could be reduced by Gaia stellar catalogue, the position uncertainty of Apophis can be significantly reduced. In particular, if only two observations (first and last) are reduced again with the Gaia catalogue, then the orbital accuracy is five times better than with the current precision of observations. 
By reducing ten (first five and last five) or twenty (first ten and last ten) optical observations of Apophis, the quality of the orbit will also be improved. 

The Gaia stellar catalogue will be very useful for astrometry because with only few observations reduced with this catalogue, the quality of the orbit will be greatly improved. For asteroids and in particular PHAs, it will help to considerably refine the orbit and the impact probability during close approaches.

\section{Conclusion}
Orbital uncertainty parameters can be useful for recovering asteroid, instrument calibration and scientific applications. They are also important in order to estimate the conditions of asteroid close approaches from Earth and eventualy impact probability.  

In this study, we particularly dealt with the Current Ephemeris Uncertainty (CEU) provided by the \textsf{astorb} database and U parameter from the \textsf{mpcorb} database. In practice, the CEU shows some advantages because it is quickly computable, precise and provides a physical value (an angle). 
Relations between CEU, $H$ magnitude, orbital arc and dynamical group have been identified. In particular, empirical linear relation between the CEU and the orbital arc have been highlighted. The CEU significantly increases while the orbital arc increases when orbital arc is between 10 and 250 days and more relatively for 250-8000 day arcs. For orbital arc larger than 8000 days, the CEU is not much improved and reaches about 0.1-0.2 arcsec. For short arc asteroids (less than 10 days), the CEU is sometimes hard to compute and remains large while the orbital arc increases.

Finally, we present statistical data about astrometric measurements. Most of asteroid observations (about 94\%) come from CCD frames. Presently, the more accurate data are radar measurements with an accuracy of about 3.3 km in distance and 6.7 km/s in velocity. Unfortunately, these measurements are still rare and mostly limited to PHAs (during close approaches). CCD frames remain relatively accurate (0.4 arcsec) and are quite numerous. These measurements provide the astrometry for most of asteroids. The bias in stellar catalogues remains the main cause of uncertainty. 

Nowadays, there are two ways for improving orbit uncertainty. One is to increase the length of orbital arc. We have shown how fast CEU decreases while the orbital arc increases. In this context, it is necessary to regularly provide asteroid observations by performing new observations or dataming. The second way is to improve the accuracy of measurements. Radar measurements appear as the most accurate but they remain few in number.  

The Gaia mission will be a revolution for asteroid dynamic and astrometry. The satellite will provide several tens of observations for most of asteroids. For PHAs, in particular the asteroid Apophis, we have shown that Gaia observations will help to improve orbit accuracy and to refine its impact probability during close approaches. In the B-plane, Gaia will decrease the uncertainty at the kilometer level instead of 10 km level currently. Moreover, with the Gaia stellar catalogue, it will be possible to reduce future and old observations with a very high level of accuracy (about 10 mas). Even if only few of the old observations could be reduced with this catalogue, the orbit quality will be about five times better than currently. In the extreme case where all observations can be reduced, the orbit uncertainty will be about 50 times smaller than currently.

\begin{acknowledgements}
This study was partly supported by contract ESA/22885/09/F/MOS/100409-R from European Space Agency. 
\end{acknowledgements}

\bibliographystyle{aa} 
\bibliography{biblio} 

\begin{thebibliography}{36}
\expandafter\ifx\csname natexlab\endcsname\relax\def\natexlab#1{#1}\fi

\bibitem[{{AstDyS}(2012)}]{astdys}
{AstDyS}. 2012, {Asteroids Dynamics Site},
  \url{http://hamilton.dm.unipi.it/astdys}

\bibitem[{{Bancelin} {et~al.}(2012{\natexlab{a}}){Bancelin}, {Colas},
  {Thuillot}, {Hestroffer}, \& {Assafin}}]{bancelin12a}
{Bancelin}, D., {Colas}, F., {Thuillot}, W., {Hestroffer}, D., \& {Assafin}, M.
  2012{\natexlab{a}}, \aap

\bibitem[{{Bancelin} {et~al.}(2012{\natexlab{b}}){Bancelin}, {Hestroffer}, \&
  {Thuillot}}]{Bancelin2012}
{Bancelin}, D., {Hestroffer}, D., \& {Thuillot}, W. 2012{\natexlab{b}},
  \planss, 73, 21

\bibitem[{{Bancelin} {et~al.}(2012{\natexlab{c}}){Bancelin}, {Hestroffer}, \&
  {Thuillot}}]{bancelin12}
{Bancelin}, D., {Hestroffer}, D., \& {Thuillot}, W. 2012{\natexlab{c}},
  Celestial Mechanics and Dynamical Astronomy, 112, 221

\bibitem[{{Bowell}(2012)}]{astorb}
{Bowell}, E. 2012, {The asteroid orbital elements database},
  \url{ftp://ftp.lowell.edu/pub/elgb/astorb.html}

\bibitem[{{Brouwer}(1937)}]{Brouwer1937}
{Brouwer}, D. 1937, Transactions of the Astronomical Observatory of Yale
  University, 6, 173

\bibitem[{{Celletti} \& {Pinzari}(2005)}]{Celletti2005}
{Celletti}, A. \& {Pinzari}, G. 2005, Celestial Mechanics and Dynamical
  Astronomy, 93, 1

\bibitem[{{Chesley}(2006)}]{chesley06}
{Chesley}, S.~R. 2006, in IAU Symposium, Vol. 229, Asteroids, Comets, Meteors,
  ed. L.~{Daniela}, M.~{Sylvio Ferraz}, \& F.~J. {Angel}, 215--228

\bibitem[{{Desmars} {et~al.}(2009){Desmars}, {Arlot}, {Arlot}, {Lainey}, \&
  {Vienne}}]{Desmars09}
{Desmars}, J., {Arlot}, S., {Arlot}, J., {Lainey}, V., \& {Vienne}, A. 2009,
  \aap, 499, 321

\bibitem[{{Dubyago}(1961)}]{Dubyago1961}
{Dubyago}, A.~D. 1961, {The Determination of Orbits} (New York: The Macmillan
  Company, 1961)

\bibitem[{{Elliot} {et~al.}(2005){Elliot}, {Kern}, {Clancy}, {Gulbis},
  {Millis}, {Buie}, {Wasserman}, {Chiang}, {Jordan}, {Trilling}, \&
  {Meech}}]{Elliot2005}
{Elliot}, J.~L., {Kern}, S.~D., {Clancy}, K.~B., {et~al.} 2005, \aj, 129, 1117

\bibitem[{{Escobal}(1965)}]{Escobal1965}
{Escobal}, P.~R. 1965, {Methods of orbit determination} (New York: Wiley)

\bibitem[{{Gauss}(1864)}]{Gauss1864}
{Gauss}, C.~F. 1864, {{Th\'eorie du mouvement des corps c\'elestes parcourant
  des sections coniques autour du soleil / trad. du "Theoria motus" de Gauss ;
  suivie de notes, par Edmond Dubois}} (Paris : A. Bertrand, 1855; trad. E.-P.
  Dubois; in 8.), 380p.

\bibitem[{{Gauss}(1809)}]{Gauss1809}
{Gauss}, K.~F. 1809, {Theoria motvs corporvm coelestivm in sectionibvs conicis
  solem ambientivm.} (Hambvrgi, Svmtibvs F.~Perthes et I.~H.~Besser, 1809.)

\bibitem[{{Herget}(1948)}]{Herget1948}
{Herget}, P. 1948, {The computation of orbits} (Lithoprinted from copy by
  Edwards Brothers, Inc., Michigan)

\bibitem[{{Hestroffer} {et~al.}(1998){Hestroffer}, {Morando}, {Hog},
  {Kovalevsky}, {Lindegren}, \& {Mignard}}]{Hestroffer1998}
{Hestroffer}, D., {Morando}, B., {Hog}, E., {et~al.} 1998, \aap, 334, 325

\bibitem[{{Jones} {et~al.}(2006){Jones}, {Gladman}, {Petit}, {Rousselot},
  {Mousis}, {Kavelaars}, {Campo Bagatin}, {Bernabeu}, {Benavidez}, {Parker},
  {Nicholson}, {Holman}, {Grav}, {Doressoundiram}, {Veillet}, {Scholl}, \&
  {Mars}}]{Jones2006}
{Jones}, R.~L., {Gladman}, B., {Petit}, J.-M., {et~al.} 2006, \icarus, 185, 508

\bibitem[{{Kochetova} {et~al.}(2009){Kochetova}, {Chernetenko}, \&
  {Shor}}]{kochetova09}
{Kochetova}, O.~M., {Chernetenko}, Y.~A., \& {Shor}, V.~A. 2009, Solar System
  Research, 43, 324

\bibitem[{{Leveau}(1880)}]{Leveau1880}
{Leveau}, G. 1880, Annales de l'Observatoire de Paris, 15, A1

\bibitem[{{Mainzer} {et~al.}(2011){Mainzer}, {Bauer}, {Grav}, {Masiero},
  {Cutri}, {Dailey}, {Eisenhardt}, {McMillan}, {Wright}, {Walker}, {Jedicke},
  {Spahr}, {Tholen}, {Alles}, {Beck}, {Brandenburg}, {Conrow}, {Evans},
  {Fowler}, {Jarrett}, {Marsh}, {Masci}, {McCallon}, {Wheelock}, {Wittman},
  {Wyatt}, {DeBaun}, {Elliott}, {Elsbury}, {Gautier}, {Gomillion}, {Leisawitz},
  {Maleszewski}, {Micheli}, \& {Wilkins}}]{Mainzer2011}
{Mainzer}, A., {Bauer}, J., {Grav}, T., {et~al.} 2011, \apj, 731, 53

\bibitem[{{Mignard} {et~al.}(2007){Mignard}, {Cellino}, {Muinonen}, {Tanga},
  {Delb{\`o}}, {Dell'Oro}, {Granvik}, {Hestroffer}, {Mouret}, {Thuillot}, \&
  {Virtanen}}]{mignard07}
{Mignard}, F., {Cellino}, A., {Muinonen}, K., {et~al.} 2007, Earth Moon and
  Planets, 101, 97

\bibitem[{{Milani} \& {Gronchi}(2010)}]{Milani2010}
{Milani}, A. \& {Gronchi}, G.~F. 2010, {Theory of Orbital Determination}
  (Cambridge University Press)

\bibitem[{{Minor Planet Center}(2012{\natexlab{a}})}]{MPC_CATOBS}
{Minor Planet Center}. 2012{\natexlab{a}}, {MPCAT-OBS: Observation Archive}, \\
  \url{http://www.minorplanetcenter.net/iau/ECS/MPCAT-OBS/MPCAT-OBS.html}

\bibitem[{{Minor Planet Center}(2012{\natexlab{b}})}]{mpcorb}
{Minor Planet Center}. 2012{\natexlab{b}}, {MPCORB database},
  \url{http://minorplanetcenter.net/iau/MPCORB.html}

\bibitem[{{Minor Planet Center}(2012{\natexlab{c}})}]{MPC_mag_size}
{Minor Planet Center}. 2012{\natexlab{c}}, {Relation between size, albedo and
  magnitude}, \url{http://www.minorplanetcenter.org/iau/lists/Sizes.html}

\bibitem[{{Minor Planet Center}(2012{\natexlab{d}})}]{MPC_U}
{Minor Planet Center}. 2012{\natexlab{d}}, {Uncertainty parameter U}, \\
  \url{http://www.minorplanetcenter.net/iau/info/UValue.html}

\bibitem[{Morando(1965)}]{Morando1965}
Morando, B. 1965, Bulletin astronomique, 1, 331

\bibitem[{{Muinonen} \& {Bowell}(1993)}]{MB93}
{Muinonen}, K. \& {Bowell}, E. 1993, Icarus, 104, 255

\bibitem[{{Muinonen} {et~al.}(2006){Muinonen}, {Virtanen}, {Granvik}, \&
  {Laakso}}]{muinonen06}
{Muinonen}, K., {Virtanen}, J., {Granvik}, M., \& {Laakso}, T. 2006, \mnras,
  368, 809

\bibitem[{{Poincar{\'e}}(1906)}]{Poincare1906}
{Poincar{\'e}}, H. 1906, Bulletin Astronomique, Serie I, 23, 161

\bibitem[{{Roy}(2005)}]{Roy2005}
{Roy}, A.~E. 2005, {Orbital motion}, fourth edition edn. (IoP, Bristol and
  Philadelphia)

\bibitem[{{Sansaturio} \& {Arratia}(2008)}]{sansaturion08}
{Sansaturio}, M.~E. \& {Arratia}, O. 2008, Earth Moon and Planets, 102, 425

\bibitem[{{Tisserand} \& {Perchot}(1899)}]{Tisserand1899}
{Tisserand}, F. \& {Perchot}, J. 1899, {Le\c{c}ons sur la d\'etermination des
  orbites} (Gauthier-Villars, Paris), 124p.

\bibitem[{{Valsecchi} {et~al.}(2003){Valsecchi}, {Milani}, {Gronchi}, \&
  {Chesley}}]{Valsecchi2003}
{Valsecchi}, G.~B., {Milani}, A., {Gronchi}, G.~F., \& {Chesley}, S.~R. 2003,
  \aap, 408, 1179

\bibitem[{{Virtanen} \& {Muinonen}(2006)}]{virtanen06}
{Virtanen}, J. \& {Muinonen}, K. 2006, \icarus, 184, 289

\bibitem[{{Yeomans} {et~al.}(1987){Yeomans}, {Ostro}, \& {Chodas}}]{Y87}
{Yeomans}, D.~K., {Ostro}, S.~J., \& {Chodas}, P.~W. 1987, \aj, 94, 189

\end{thebibliography}

\end{document}